\documentclass[prb,aps,twocolumn,nofootinbib,superscriptaddress,floatfix]{revtex4}

\pdfoutput=1 

\usepackage{soul}

\usepackage[english]{babel}
\usepackage[utf8]{inputenc}
\usepackage[T1]{fontenc}

\usepackage{hyperref}
\usepackage{exscale,times}
\usepackage{graphicx,color}
\usepackage{latexsym}
\usepackage{epsfig}
\usepackage{dsfont}
\usepackage{amsbsy,bm,amssymb,amsmath}

\newcommand{\bra}[1]{\langle#1|}
\newcommand{\ket}[1]{|#1\rangle}

\newcommand{\ketbra}[2]{\vert {#1} \rangle \langle{#2}\vert}

\begin{document}

\title{Experimental on-demand recovery of entanglement by local operations within non-Markovian dynamics}

\author{Adeline Orieux}
\affiliation{Dipartimento di Fisica, Sapienza Universit\`a di Roma, Piazzale Aldo Moro, 5, I-00185 Roma, Italy}

\author{Antonio d'Arrigo}
\affiliation{CNR-IMM UOS Universit\`a (MATIS), Consiglio Nazionale delle Ricerche, Via Santa Sofia 64, 95123 Catania, Italy}
\affiliation{Dipartimento di Fisica e Astronomia, Universit\`a degli Studi Catania, Via Santa Sofia 64, 95123 Catania, Italy}

\author{Giacomo Ferranti}
\affiliation{Dipartimento di Fisica, Sapienza Universit\`a di Roma, Piazzale Aldo Moro, 5, I-00185 Roma, Italy}

\author{Rosario Lo Franco}
\affiliation{Dipartimento di Fisica e Chimica, Universit\`a di Palermo, via Archirafi 36, 90123 Palermo, Italy}
\affiliation{Instituto de F{\'{i}}sica de S{\~{a}}o Carlos, Universidade de S{\~{a}}o Paulo, Caixa Postal 369, 13560-970 S{\~{a}}o Carlos, S{\~{a}}o Paulo, Brazil}
\affiliation{School of Mathematical Sciences, The University of Nottingham, University Park, Nottingham NG7 2RD, United Kingdom.}

\author{Giuliano Benenti}
\affiliation{Center for Nonlinear and Complex Systems, Universit\`a degli Studi dell'Insubria, Via Valleggio 11, 22100 Como, Italy}
\affiliation{Istituto Nazionale di Fisica Nucleare, Sezione di Milano, via Celoria 16, 20133 Milano, Italy}

\author{Elisabetta Paladino}
\affiliation{Dipartimento di Fisica e Astronomia, Universit\`a degli Studi Catania, Via Santa Sofia 64, 95123 Catania, Italy}
\affiliation{CNR-IMM UOS Universit\`a (MATIS), Consiglio Nazionale delle Ricerche, Via Santa Sofia 64, 95123 Catania, Italy}
\affiliation{Istituto Nazionale di Fisica Nucleare, Sezione di Catania, Via S. Sofia 64, 95123 Catania, Italy}

\author{Giuseppe Falci}
\affiliation{Dipartimento di Fisica e Astronomia, Universit\`a degli Studi Catania, Via Santa Sofia 64, 95123 Catania, Italy}
\affiliation{CNR-IMM UOS Universit\`a (MATIS), Consiglio Nazionale delle Ricerche, Via Santa Sofia 64, 95123 Catania, Italy}
\affiliation{Istituto Nazionale di Fisica Nucleare, Sezione di Catania, Via S. Sofia 64, 95123 Catania, Italy}

\author{Fabio Sciarrino}
\affiliation{Dipartimento di Fisica, Sapienza Universit\`a di Roma, Piazzale Aldo Moro, 5, I-00185 Roma, Italy}

\author{Paolo Mataloni}
\affiliation{Dipartimento di Fisica, Sapienza Universit\`a di Roma, Piazzale Aldo Moro, 5, I-00185 Roma, Italy}

\date{\today}

\begin{abstract}
In many applications entanglement must be distributed through noisy communication channels that unavoidably degrade it. Entanglement cannot be generated by local operations and classical communication (LOCC), implying that once it has been distributed it is not possible to recreate it by LOCC.
Recovery of entanglement by purely local control is however not forbidden in the presence of non-Markovian dynamics, and here we demonstrate in two all-optical experiments that such entanglement restoration can even be achieved on-demand.
First, we implement an open-loop control scheme based on a purely local operation, without acquiring any information on the environment; then, we use a closed-loop scheme in which the environment is measured, the outcome controling the local operations on the system. The restored entanglement is a manifestation of "hidden" quantum correlations resumed by the local control. Relying on local control, both schemes improve the efficiency of entanglement sharing in distributed quantum networks.
\end{abstract}

\maketitle

Quantum mechanics promises breakthroughs in computing and information fields, some of which are already available~\cite{lunghi2013,ekert2014}. Similarly to networking in classical information and computation, the advent of quantum networks envisages further advancements in information science~\cite{kimble2008,sciarrino2012,perseguers2013}. Tasks such as measurements, computing and memorization may be performed by subsystems implemented on different platforms~\cite{hybrid}, networking also providing the large amount of resources required to fault tolerant computation schemes~\cite{steane1998,preskill1998}.
Recent experimental up-scaling of quantum processors made it clear that for hardware-intrinsic noise sources the low-decoherence DiVincenzo criterion~\cite{divincenzo2000} could be met by subsystems of limited size~\cite{ladd2010} in distributed architectures. This is a new, and uniquely "quantum" feature of networking.

The roadmap towards distributed networks critically relies on the possibility that pairs (or clusters) of nodes share entanglement~\cite{PlenioReview,HorodeckiReview}. This is the fundamental resource allowing remote quantum teleportation~\cite{bennett1993,zeilinger1997,boschi1998,ursin2007} of an unknown quantum state or secure keys distribution for cryptographic purposes~\cite{gisin2002}.
In particular for the scope of our work it is important to note that entanglement could enable universal quantum computation in networks of noninteracting nodes, provided single qubit local operations are possible~\cite{gottesman1999}. Sharing entanglement can reduce the communication complexity, that is the minimal information exchange required to solve a given problem distributed among separated parts~\cite{bruckner2004}. Very recently, it has been proposed that entanglement may improve accuracy and precision in applications related to global positioning and timing by networking geographically remote atomic clocks~\cite{komer2014}.

In order to use entanglement as a non-local resource~\cite{PlenioReview,HorodeckiReview}, it must be generated somewhere and then distributed amongst different parties. However noise unavoidably affects distribution and storage of entanglement, determining its degradation.
It is well known that different parties cannot create any further entanglement, if they are only allowed to operate locally, i.e. on their own subsystem, exchanging at most classical information~\cite{bennett1996}. 

Our scope is distributed networks of spatially separated quantum nodes, each subject to a local environment. They model the structure of physically relevant distributed architectures. Remarkably, when environments induce a system dynamics which can be physically unraveled into an ensemble of entangled pure state evolutions, it may happen that quantum correlations of initially entangled states are not destroyed. Indeed even if entanglement appears degraded when measured on the averaged state, it is not lost but rather \textit{hidden} in the lack of classical knowledge about the elements of the ensemble~\cite{darrigo2014Annals}.
In this case, leveraging the existence of such classical information, the initially shared entanglement can be restored at an arbitrary time without resorting to non-local operations.
 
An operational scenario emerges in which local controls can be used for on-demand restoration of entanglement in distributed architectures. Here we demonstrate this concept by two all-optical experiments. We consider two entangled photons subject to local environments as an instance of distributed entanglement.
In the first experiment a local environment produces low-frequency noise~\cite{PaladinoReview2013}, and it is shown how by \textit{local open loop control}~\cite{vandersypen2005} the initial entanglement is recovered, even if in the absence of control it would be degraded to very low values.
In the second experiment, decoherence is due to the coupling between a subsystem (photon) and a quantum environment, whose degrees of freedom are experimentally accessible. A measurement on this environment and a subsequent conditional local operation on the system implement a \textit{local closed-loop control}~\cite{wiseman2010book}, succeeding in restoring the initial entanglement. The relevance of these setups to different physical systems is discussed at the end of this work.

The non-Markovian nature of the dynamics~\cite{cohen-Tannoudji,rivas2014} is a key ingredient for the entanglement recovery in our experiments. Certain physical aspects are common to other phenomena involving non-Markovianity, such as spontaneous entanglement revivals during the system dynamics~\cite{bellomo2007PRL,liu_natphys_2011,chiu2012,lofranco2012PRA,xu_natcomm_2013}, or entanglement preservation by dynamical decoupling~\cite{wang2011,roy_pra_2011,loFranco2014}.
Here, at variance vith these examples, we engineer the overal dynamics by a local control, and we demonstrate that distributed entanglement can be fully restored on-demand by suitable local operations even though it would vanish in the absence of active control.
\section*{Results}

We consider a prototype distributed quantum network implemented by an all-optical setup where the information carriers are two photons, $A$ and $B$. The information is encoded in the photon polarization, being either horizontal $|H \rangle$, or vertical $|V \rangle$. The system $AB$ is prepared in the two-photon entangled Bell state $\ket{\Psi^-}=\big(\ket{HV}-\ket{VH}\big)/\sqrt{2}$. Qubits $A$ and $B$ propagate in free-space (communication channel) experiencing {\em local} interactions with different environmental degrees of freedom.

\subsection{Theoretical framework}
Typically the information available at time $t$ on the system $AB$ is encoded in the reduced density matrix $\rho(t)$, obtained after tracing over the environment. It can be decomposed, in an infinite number of ways, in terms of pure states $\ket{\nu(t)}$, each one occurring with a probability$p_\nu(t)$: $\rho(t)=\sum_\nu p_\nu(t) \ketbra{\nu(t)}{\nu(t)}$. The entanglement of the average state $\rho(t)$ is
\begin{equation}
E_\rho(t)=E\Big(\sum_\nu p_\nu(t) \ketbra{\nu(t)}{\nu(t)}\Big),
\label{eq-results:entanglement-of-averaged-state}
\end{equation}
where $E$ is some measure, reducing for pure states to the entropy of entanglement~\cite{PlenioReview,HorodeckiReview}.
In this work we study the entanglement of formation~\cite{PlenioReview,HorodeckiReview} $E_f$, but we discuss our results in terms of the concurrence $C$, an entanglement measure with a simpler and very intuitive form~\cite{wootters1998PRL}. Note that $E_f$ monotonically depends on the concurrence, and can be readily calculated from it as $E_f(C)$, see Appendix.

In the absence of interaction with the environment, the density matrix does not evolve $\rho(t)=\rho(0)=\ket{\Psi^-}\bra{\Psi^-}$, whereas due to the interaction of the qubits with the environment, $\rho(t)$ evolves in a statistical mixture. The corresponding entanglement quantifier is no longer equal to that of the initial state $\ket{\Psi^-}$.

Here we demonstrate that, acting by suitable {\em local} controls, the entanglement initially present in the system $AB$ can be restored. This is possible if $\rho(t)$ corresponds to a specific physical decomposition $\bar{\cal Q}(t) \equiv \{(p_{\bar\nu(t)}, \ket{\bar \nu(t)})\}$, and we are somehow able to tag each member of the ensemble $\bar{\cal Q}(t)$ and to know its state. 
In this case we can distill the average entanglement of $\bar{\cal Q}$
\begin{equation}
E_{\bar{\cal Q}}(t)=\sum_{\bar\nu} p_{\bar\nu}(t) E\Big(\ketbra{\bar\nu_{}(t)}{\bar\nu_{}(t)}\Big),
\label{eq-results:averaged-entanglement-of-eachstate}
\end{equation}
by using {\em local} operations and classical communication only.
For any convex measure $E$ of entanglement, we have that $E_{\bar{\cal Q}}\ge E_{\rho}$. The inequality has a natural meaning, namely the classical knowledge of the state of each member of the quantum ensemble $\bar{\cal Q}(t)$ allows for a larger entanglement with respect to situations where this information is not available.

\subsection{The experiments}
In the two experiments we present here, the two-photon state $AB$ is generated by a spontaneous parametric down-conversion (SPDC) source of photon pairs based on~\cite{Kwiat1995}. In both experiments, the degraded entanglement $E_\rho(t)$, Eq.(\ref{eq-results:entanglement-of-averaged-state}), is restored by a local control, whose effect is to make available the entanglement $E_{\bar{\cal Q}}(t)$, Eq.(\ref{eq-results:averaged-entanglement-of-eachstate}). For the sake of simplicity photon $A$ is measured directly whereas photon $B$ is sent through a noisy channel before being measured. We address situations where the evolution of the system $AB$ is non-unitary, due to the interaction with a local environment $O$ inducing a non-Markovian dynamics of $AB$~\cite{rivas2010,chiu2012}. 
 
In a first experiment we simulate classical non-Markovian noise through a sequence of liquid crystal retarders which add random phases to photon $B$ during propagation (pure dephasing). As a result, the initial entanglement of the system $AB$  decays monotonically as a function of the channel length. Despite we are aware of the physical decomposition in terms of pure states $\ket{\bar{\nu}(t)}$ and probabilities $p_{\bar{\nu}}(t)$, we are not able to tag each state $\bar{\nu}$ in the ensemble. 
Therefore the entanglement of the average state is given by  Eq. (\ref{eq-results:entanglement-of-averaged-state}). The action of a bit-flip on photon $B$ applied at half-way propagation along the channel restores the initial entanglement (\textit{open-loop control}). Therefore such a {\em local} control is able to retrieve the classical information on each member of the quantum ensemble $\bar{\cal Q}(t)$, thus allowing to make available the average entanglement $E_{\bar{\cal Q}}$, Eq.~(\ref{eq-results:averaged-entanglement-of-eachstate}). 
The amount of entanglement recovered by this technique depends on the degree of correlations (non-Markovianity) among the environment-added random phases.

In a second experiment, we simulate a quantum environment by a third qubit, $O$, interacting with qubit $B$ through a controlled-NOT (CNOT) gate. The coherent exchange of information between $AB$ and the environment $O$ roots the emergence of memory effects and of a non-Markovian evolution of the reduced system $AB$. 
Measurement of the environment in a given basis makes it possible the physical selection of a given quantum ensemble $\bar{\cal Q}(t)$, allowing at the same time to tag each state of this ensemble. The average entanglement of $\bar{\cal Q}(t)$ (which can be equal to the initial entanglement) is then obtained, by applying on qubit $B$ a {\em local} operation on qubit $B$, which depends on the (classical) information gained on the environmental state (\textit{closed-loop control}). 

Both experiments and their results are described in detail in the following subsections.

\subsection{Open-loop control}
\begin{figure}[h]
\centering
\includegraphics[width=\columnwidth]{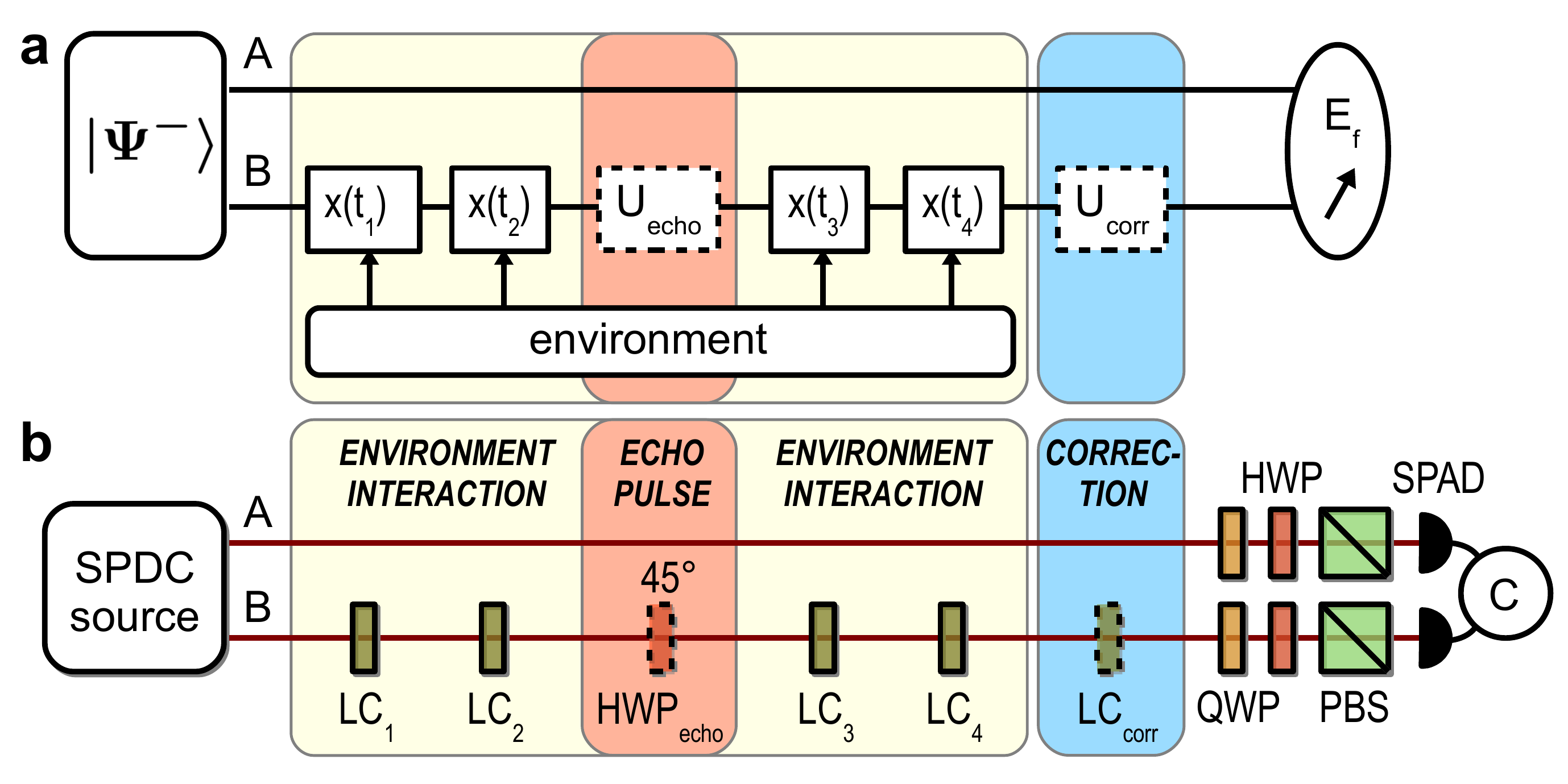}
\caption{{\bf Open-loop set-up.} {\bf a} Conceptual scheme. Qubits A and B are prepared in the Bell state $\ket{\Psi^-}$. Qubit B interacts stroboscopically with the environment through the 4 random phases $x(t_k)$. The noise 
induced by the environment can be compensated either with a rephasing unitary $U_{corr}$ or with an echo-pulse unitary $U_{echo}=\sigma_x$.  $E_f$: entanglement of formation measurement. {\bf b} Experimental implementation. LC: liquid cristal retarder, HWP: half-wave plate, QWP: quarter-wave plate, PBS: polarizing beam-splitter, SPAD: single photon avalanche photodiode, C: coincidence counting electronics.}
\label{Fig_setup_open}
\end{figure}

In this experiment, the photon $B$ interacts with a classical environment $O$ described by a stochastic process $x(t)$. The corresponding noisy channel acting on $B$ is designed to induce pure dephasing, according to the Hamiltonian:
${\cal H}_B(t)=x(t)\delta(t-t_k)\sigma_{z}/2$,
where $\sigma_z=\ketbra{H}{H}-\ketbra{V}{V}$ and $\delta$ is the Dirac delta function. Here, the interaction between $B$ and $O$ takes place stroboscopically at times $t_k$ (see Fig.~\ref{Fig_setup_open}~a).
This Hamiltonian is experimentally realized by sending the photon through a sequence of four liquid crystal retarders (LC$_k$), each one introducing a phase $\chi_k\equiv x(t_k)$ between the photon polarization components: $\alpha\ket{H_B}+\beta\ket{V_B}\to\alpha\ket{H_B}+e^{i\chi_k}\beta\ket{V_B}$. The phase $\chi_k$ depends on the voltage $V_k$ applied to LC$_k$ and can be varied continuously from 0 to $\pi$ (see Fig.~\ref{Fig_setup_open}~b).
To simulate the stochastic process, we generate a set of $N$ random phase sequences $\vec{\chi}=\{\chi_1,\chi_2,\chi_3,\chi_4\}$, in which the phases $\chi_k$ are Gaussian random variables with the same variance $\sigma^2$, and correlations $\mu\equiv\langle \chi_k\,\chi_{k+1}\rangle/\sigma^2$, $\mu\, \in\, [0,1]$ (see Appendix). The system dynamics averaging with respect to these sequences is obtained by mixing together the tomographic measurement data obtained with each of the $N$ random phases sequences.

\noindent We investigated three different situations: a) the uncontrolled dynamics, where we simply look at the entanglement degradation resulting from the noisy channel; b) the controlled dynamics where we show how accessing classical information on the environment allows to operate corrections which fully restore the entanglement, and c) the echoed dynamics where entanglement is recovered by a simple local operation with no need to have access to the classical information. 

\begin{figure}[h]
\centering
\includegraphics[width=\columnwidth]{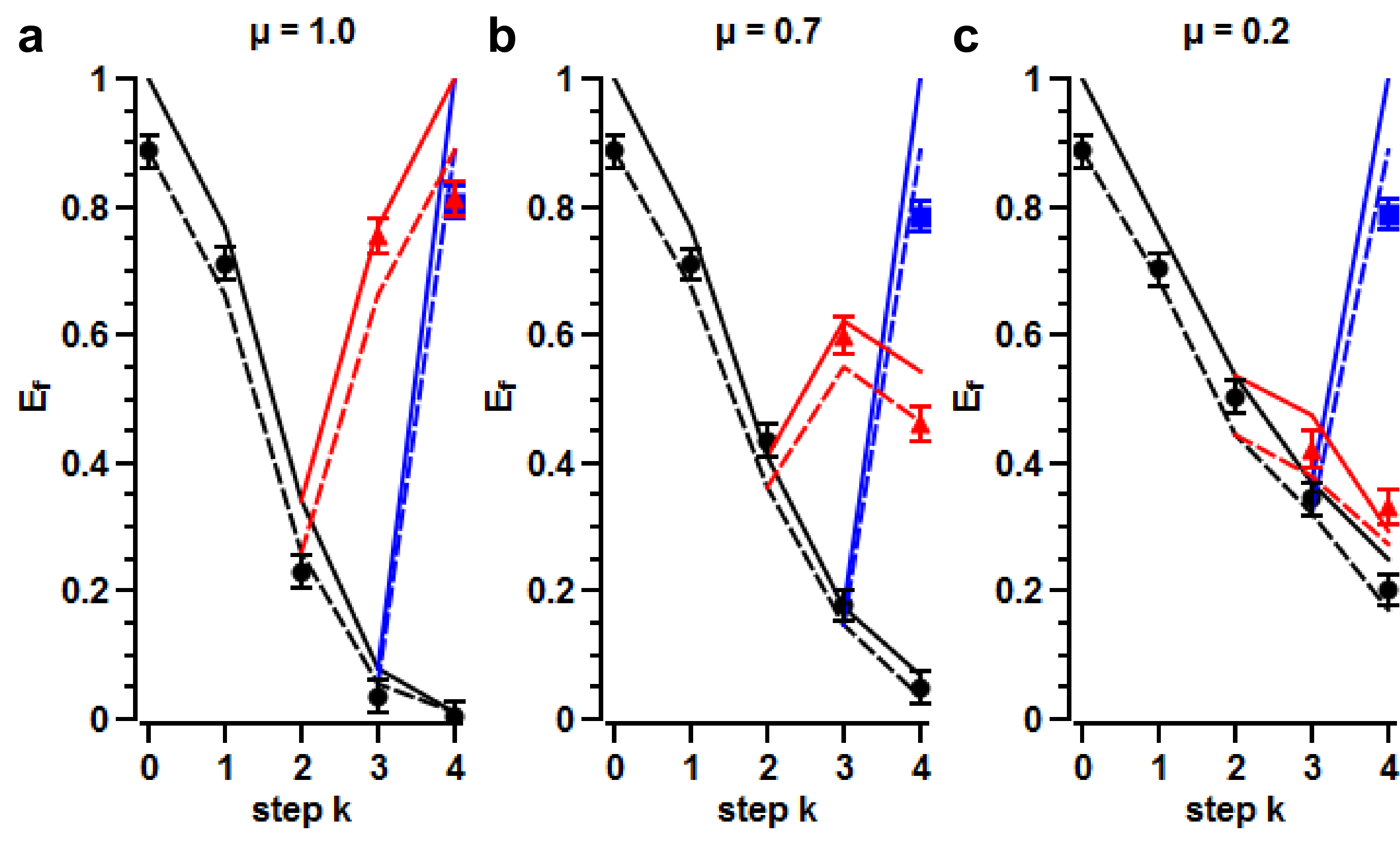}
\caption{{\bf Open-loop results.} Entanglement of formation $E_f$ measured at each step $k$ for three values of $\mu$ ({\bf a} $\mu=1.0$. {\bf b} $\mu=0.7$. {\bf c} $\mu=0.2$). Symbols: experimental data points, continuous line: theoretical calculations for a Bell state, dotted lines: simulations for a state with a fidelity $F=0.96$ to a Bell state. Black, blue and red colours 
correspond respectively to the uncontrolled, corrected and echoed dynamics. The error bars are derived from propagating the Poissonian statistical errors of the photon coincidence counting.}
\label{Fig_results_open}
\end{figure}

\textbf{Uncontrolled dynamics.--} 
The reduced dynamics of $AB$ induced by noise is in this case unambiguously described by the quantum ensemble $\bar{\cal Q}=\big\{p(\vec{\chi}),\,\ket{\bar{\nu}_{\chi_1,...,\chi_k}}\big\}$, where $p(\vec{\chi})$ stands for the joint probability $p(\chi_1,\chi_2,\chi_3,\chi_4)$ and
\begin{equation}
\ket{{\nu}_{\chi_1,...,\chi_k}}=\frac{1}{\sqrt{2}}\,(\ket{HV}-e^{i\varphi_k}\ket{VH}),
\label{eq-results:1stExp_pure-state}
\end{equation}
with $\varphi_k=\sum_{j=1}^k \chi_j$ the overall phase accumulated up to step $k$. As a result, each state of the ensemble is maximally entangled, so that also the average entanglement (\ref{eq-results:averaged-entanglement-of-eachstate}) is maximum for any $k$: $E_{\bar{\cal Q}}(k)=1$. However, the entanglement (\ref{eq-results:entanglement-of-averaged-state}) of the average state, $E_{\rho}(k)$, exhibits quite a different behavior. The system concurrence for the \emph{uncontrolled} dynamics decays with $k$: in the case $\mu=1$ (full correlations), theory gives $C_{unco}(k)=e^{-\frac{1}{2}\sigma^2k^2}$. For $\mu<1$ the concurrence, though more involved (see Appendix), shows a similarly decaying behavior.
In the experiment we measured the entanglement $E_{\rho}(k)$ obtained for three different values of $\mu$ with the generated sets of random phase sequences $\vec{\chi}$. For $k<4$, the LC$_i$ with $i>k$ are set at a constant phase $\frac{\pi}{2}$ instead of $\chi_i$. The experimental (black symbols) and theoretical results (black lines) are presented in Fig.~\ref{Fig_results_open}. These results show that the system entanglement decreases as the accumulated phase $\varphi_k$ grows.

\textbf{Corrected dynamics.--} 
In this case, we know the induced noise $\vec{\chi}$ (since we generate it), and we can compensate it. Indeed, all that is needed is to insert another LC (LC$_{corr}$) after the channel and to apply to it a voltage so as to produce a phase $x_{corr}=-\varphi_4$, see Fig.~\ref{Fig_setup_open}. In practice, having only four LCs available, we used LC$_4$ as the correction step and set it at a phase $x_4=-\varphi_3$. For such \emph{corrected} dynamics the initial state is fully recovered, together with its entanglement: $C_{corr}(4)=1$, for any $\mu$, see blue symbols (experiment) and blue lines (theory) in Fig.~\ref{Fig_results_open}. 
This demonstrates that the channel-induced degradation of entanglement is only due to a lack of classical knowledge: once $\varphi_3$ is known, entanglement is recovered by a local phase-shift operation.

\textbf{Echoed dynamics} (open-loop control)\textbf{.--} 
Accessing information on the environment is not possible in practice in real quantum networks, where noise sources correspond to a large number of uncontrollable environmental degrees of freedom. Nevertheless, an open loop scheme operated by local control may still allow to recover the entanglement. 
To demonstrate this we use a control technique introduced in NMR~\cite{Schlichter92}: just after the step $k=2$, we apply a local bit-flip operation ($U_{echo}=\sigma_{x}$) on photon $B$, which flips its polarization: $\sigma_{x}\ket{H}=\ket{V}$ and $\sigma_{x}\ket{V}=\ket{H}$, see Fig.~\ref{Fig_setup_open}~a. Experimentally we insert a half-wave plate at $45^o$ between LC$_2$ and LC$_3$ when measuring the entanglement at $k=3,4$, see Fig.~\ref{Fig_setup_open}~b.
Our purpose is to induce an \emph{entanglement echo} in the system dynamics. The state (\ref{eq-results:1stExp_pure-state}) which describes the system in each run of the experiment for $k\ge3$ now reads: 
\begin{equation}
\ket{\tilde{\nu}_{\chi_1,...,\chi_k}}=\frac{1}{\sqrt{2}}\,(\ket{HH}-e^{i(\chi_1+\chi_2-\sum_{j=3}^k \chi_j)}\ket{VV}).
\label{eq-results:1stExp_pure-state-echoed}
\end{equation}
The local pulse is equivalent to a change of sign of the phases acquired at steps $k=3,4$, and it may tend to cancel the effect of $\varphi_2=\chi_1+\chi_2$ if the four $\chi_i$ phases are to a certain extent correlated. In general the echo pulse is expected to favor the recovery of the average state entanglement, especially for an environment with \textit{non-Markovian} correlations.
 
The plots in Fig.~\ref{Fig_results_open} show that, indeed, the entanglement for the echoed dynamics starts to increase at $k\ge3$, see red symbols (experiment) and red lines (theory). It is also clear that entanglement recovery strongly depends on correlations. For $\mu=1$ (Fig.~\ref{Fig_results_open}~a), the characteristic time scale of the environment dynamics is much larger than that of the system dynamics. In this case full entanglement recovery is possible: $\chi_3+\chi_4=-\varphi_2$ and we have that $C_{echo}(k\ge3)=e^{-\frac{1}{2}\sigma^2(k-4)^2}$ (red line), see Appendix. Partial recovery is possible when $0<\mu<1$ (Fig.~\ref{Fig_results_open}~b for $\mu=0.7$ and Fig.~\ref{Fig_results_open}~c for $\mu=0.2$). In that case, $\chi_3+\chi_4$ cancels $\varphi_2$ only partially, and we have to deal with a bit more involved expression for the concurrence (red lines), see Appendix. Let us stress that, whereas in the corrected dynamics we used the knowledge of the sequence $\vec{\chi}$ to cancel the accumulated random phase, in the echoed dynamics the (partial) cancellation of the phase needs no knowledge of the environment.

Our experimental data reported in Fig.~\ref{Fig_results_open} show a very good entanglement recovery, both for the corrected and the echoed dynamics. Deviations from the ideal expectations are mainly due to the imperfect preparation of the input state: in Fig.~\ref{Fig_results_open} we also plot (dashed lines) the expected entanglement for a mixed input state with a (measured) fidelity $F_{\Psi^-}=\bra{\Psi^-}\rho_{in}\ket{\Psi^-}=0.96$ to the ideal input state $\ket{\Psi^-}$, see Eqs.~(\ref{methods:mixed-initial-state}) and (\ref{methods:fidelity-mixed-initial-state}) in Appendix. Black and red dashed lines refer respectively to the uncontrolled and echoed entanglement, which are derived from Eqs.~(\ref{methods:Concurrence-UncontrolledDynamics-ImperfectPreparation}) and (\ref{methods:Concurrence-EchoedDynamics-ImperfectPreparation}) in Appendix. For the corrected dynamics (blue squares), the recovered entanglement at step $k=4$ does not quite reach the initial entanglement at $k=0$, this is due to the finite precision we have on setting the phases for each LC.

\subsection{Closed-loop control} \label{sec:Closed-loop control}

\begin{figure}[h]
\centering
\includegraphics[width=\columnwidth]{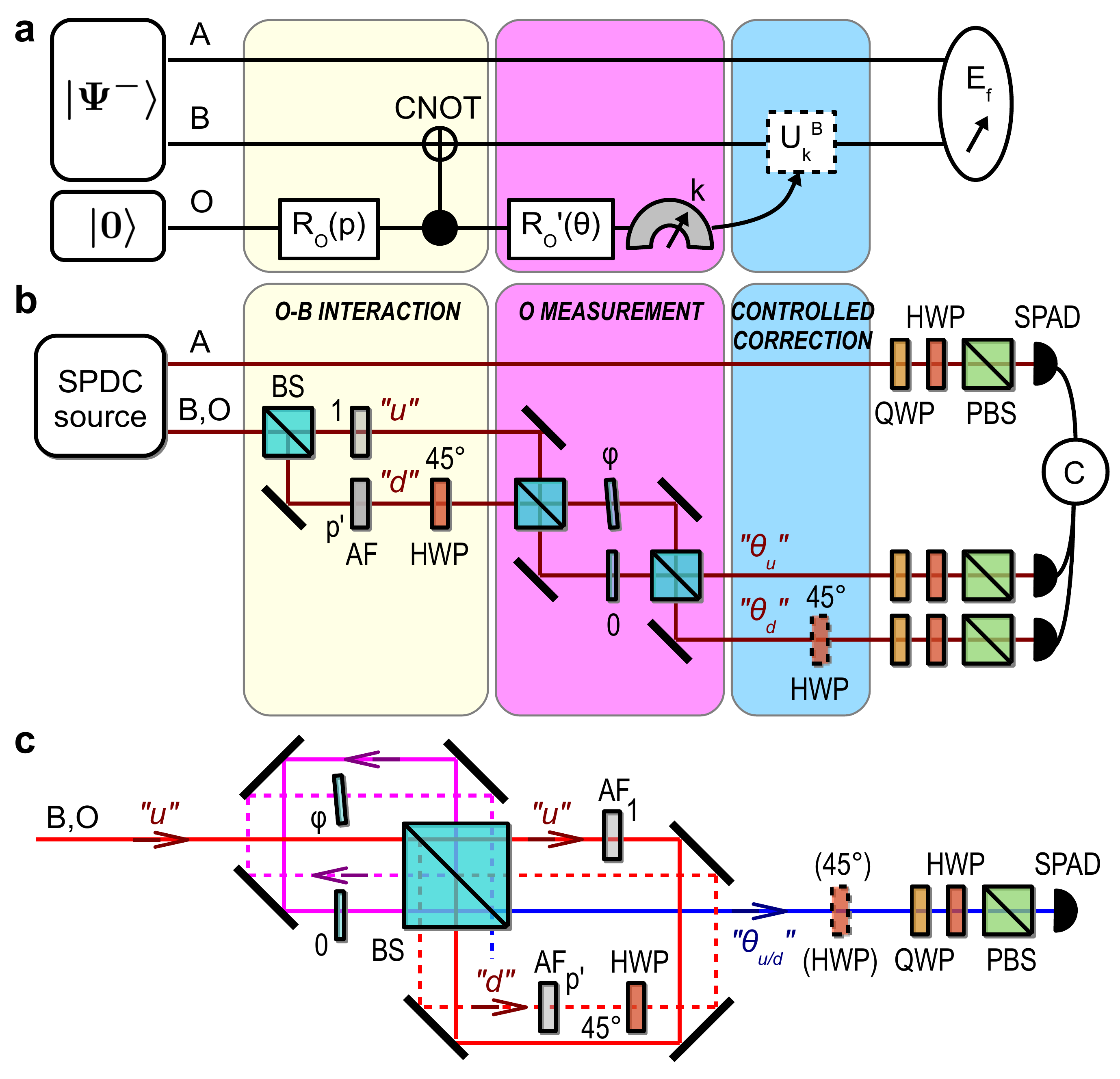}
\caption{{\bf Closed-loop set-up.} {\bf a} Conceptual scheme. Qubits A and B are prepared in the Bell state $\ket{\Psi^-}$, the environment O, initially in the state $\ket{0}$, undergoes a rotation $R_O(p)$ that splits its state between the "up" and "down" paths. O then interacts with qubit B through a controlled-NOT gate (CNOT). O is measured in a rotated basis by $R_O'(\theta)$ which gives a result k (either the "$\theta_u$" or the "$\theta_d$" path). The environment interaction can be corrected with a unitary $U_k^B$. $E_f$: entanglement of formation measurement. {\bf b} Experimental implementation. "u" and "d": "up" and "down" path states of the environment, BS: balanced beam-splitter, AF: attenuation filter, $\varphi$: thin glass plate to adjust the phase $\varphi=2\theta$, HWP: half-wave plate, QWP: quarter-wave plate, PBS: polarizing beam-splitter, SPAD: single photon avalanche photodiode, C: coincidence counting electronics. {\bf c} Actual experimental set-up for photon B, see Methods. The red paths correspond to the "O-B interaction" part, the pink to the "O measurement" and the blue to the "controlled correction" and polarization tomographic measurement. Both exit modes "$\theta_u$" and "$\theta_d$" of {\bf b} are measured successively on the same exit port "$\theta_{u/d}$" for two complementary phase settings $\varphi$ and $\pi-\varphi$; the control HWP at $45^o$ is inserted only for $\pi-\varphi$.}
\label{Fig_setup_closed}
\end{figure}

In this second experiment, qubits $A$ and $B$ are still encoded in the polarization of photons $A$ and $B$, the environment being now a third qubit $O$, encoded in the longitudinal momentum degree of freedom (the path) of photon $B$. The state of $O$ can be either "up", $|u \rangle =|0\rangle$, or "down", $|d \rangle =|1\rangle$, see Fig.~\ref{Fig_setup_closed}. The system plus environment $ABO$ is initially prepared in the state
\begin{eqnarray}
\ket{\Psi_{ABO}}=\ket{\Psi^-}\otimes\ket{u}.
\label{results:2ndExp-initial-ABO-state}
\end{eqnarray}
The interaction of $B$ with the environment $O$ is engineered as follows: $O$ is first rotated by the gate ${\cal R}_O(p)= \sqrt{1-p}\,\sigma_{z_O}+\sqrt{p}\,\sigma_{x_O}$, where $\sigma_{z_O}\equiv\ketbra{u}{u}-\ketbra{d}{d}$ and $\sigma_{x_O}\equiv\ketbra{u}{d}+\ketbra{d}{u}$; subsequently, $B$ undergoes the CNOT gate ${G}_{BO}=\openone_B\otimes\ketbra{u}{u}+\sigma_{x_B}\otimes\ketbra{d}{d}$, which may flip the polarization of $B$ according to the state of $O$, see Fig.~\ref{Fig_setup_closed}~a. 
The gate ${\cal R}_O(p)$ is experimentally implemented by  a balanced beam-splitter and an attenuation filter in both its output ports, with a ratio $p'=p/(1-p)$ between their respective intensity transmission coefficients; the CNOT gate ${G}_{BO}$ is implemented by an half-wave plate at $45^o$ in the "down" path of photon $B$, see Fig.~\ref{Fig_setup_closed}~b.

\begin{figure}[h]
\centering
\includegraphics[width=\columnwidth]{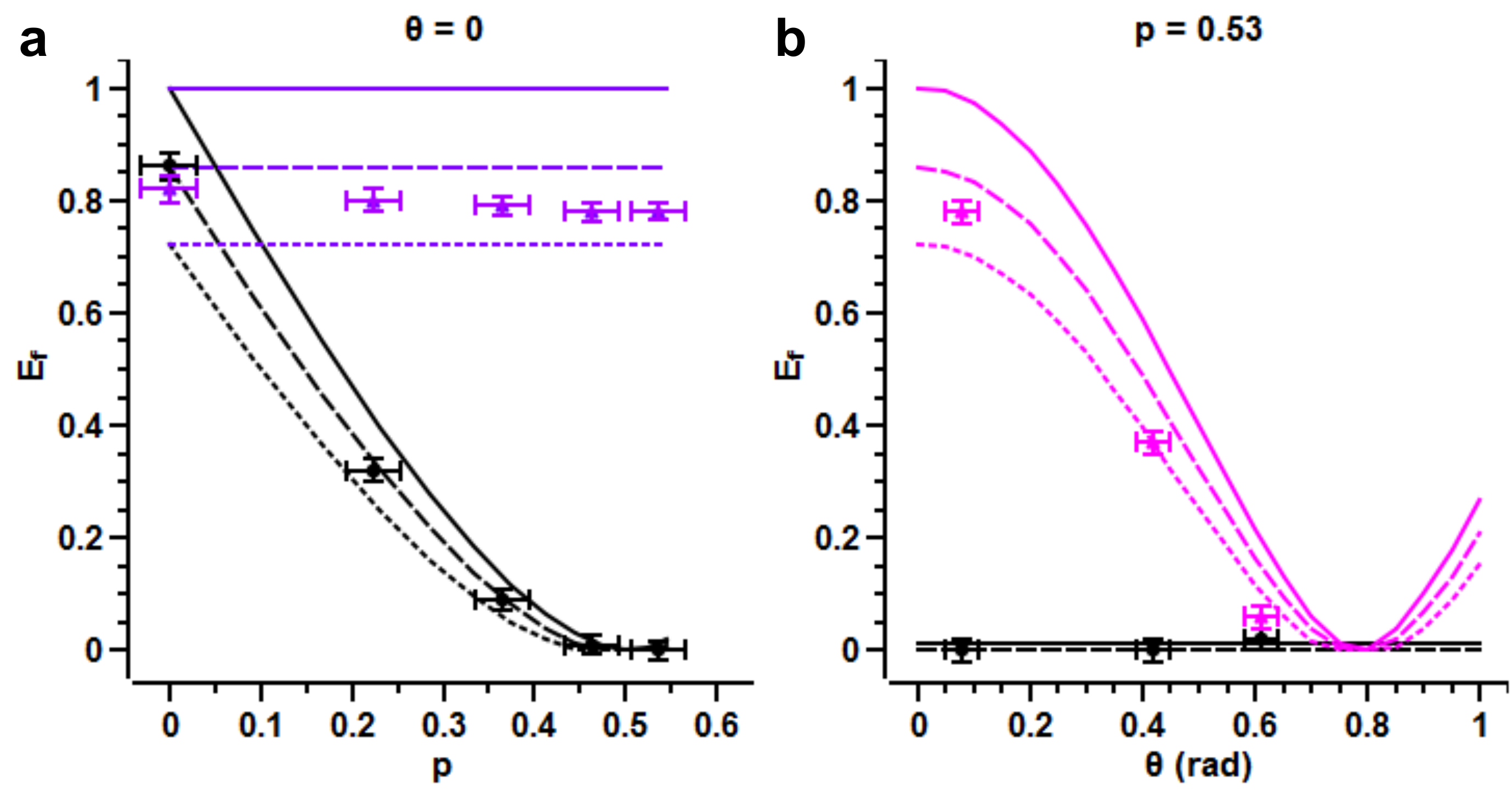}
\caption{{\bf Closed-loop results.} Entanglement of formation $E_f$ as a function of {\bf a} the environment rotation parameter $p$, {\bf b} the measurement basis angle $\theta$. Symbols: experimental data points, continuous line: theoretical calculations for a Bell state, dotted (dashed) lines: simulations for a state with a fidelity $F=0.90$ ($F=0.95$) to a Bell state. The black colour corresponds to the uncontrolled dynamics and the violet/pink colours to the corrected dynamics. The error bars determination is given in the Methods.}
\label{Fig_results_closed}
\end{figure}

\textbf{Uncontrolled dynamics.--} 
The state of $ABO$ after the ${BO}$ interaction is:
\begin{equation}
    \ket{\Psi^{(1)}_{ABO}(p)}=\sqrt{1-p}\ket{\Psi^-}\otimes\ket{u}+\sqrt{p}\ket{\Phi^-}\otimes\ket{d},
\label{eq-results:2ndExp-01}
\end{equation}
where $\ket{\Phi^-}=\big(\ket{HH}-\ket{VV}\big)/\sqrt{2}$ is another Bell state.
After tracing out the environment, the system is described by $\rho_{out}(p)=\textrm{Tr}_O\big[\ketbra{\Psi^{(1)}_{ABO}(p)}{\Psi^{(1)}_{ABO}(p)}\big]$, with concurrence $C_{unco}(p)=|1-2p|$. No further operation is applied on $B$ before we measure the entanglement of $AB$. The corresponding  entanglement of formation $E_f$ depends on $p$, exhibiting a monotonous decay in the range $[0,1/2]$, see the black curve (theory) and points (experiment) in Fig.~\ref{Fig_results_closed}~a.

\textbf{Controlled dynamics.--} 
We show how it is possible to recover entanglement by a "closed-loop" like control scheme. It consists of two steps: 1) after the gate $G_{BO}$, $O$ is projectively measured in the basis $\{\ket{\theta_u}= e^{i\theta\sigma_{x_O}}\ket{u}, \ket{\theta_d}\,=\,e^{i\theta\sigma_{x_O}}\ket{d}\}$, and 2) depending on the outcome $k$, a unitary operation $U^B_k$ is performed on $B$, $U^B_u=\openone_{B}$ and $U^B_d=\sigma_{x_B}$, see Fig.~\ref{Fig_setup_closed}~a. Experimentally, the selection of the $O$ measurement basis is obtained by a Mach-Zehnder interferometer with a variable phase $\varphi=2\theta$ which realizes the rotation ${\cal R}'_O(\theta)=e^{-i\theta\sigma_{x_O}}$, see Fig.~\ref{Fig_setup_closed}~b. The unitary $U^B_k$ is obtained by doing nothing further on the output path $\ket{\theta_u}$ and inserting a half-wave plate at $45^o$ in the path $\ket{\theta_d}$. The entanglement in $AB$ is measured by mixing the polarization tomography data obtained on both output paths. Note that we actually used a folded version of the set-up presented in Fig.~\ref{Fig_setup_closed}~b, see Appendix and Fig.~\ref{Fig_setup_closed}~c.

The fact that we can distinguish the modes $k=u,d$ emerging from the second interferometer, allows to associate on a physical basis the ensemble $\bar{\cal Q}= \big\{\big(p_{\theta_u},\ket{\psi_{\theta_u}}\big), \big(p_{\theta_d},\ket{\psi_{\theta_d}}\big)\big\}$ (see Appendix) to the state of the system.
Moreover, we tag the actual state of the ensemble during each run of the experiment. This classical information enables us to apply some suitable local control which depends on the actual state. The goal is to obtain a new quantum ensemble $\bar{\cal Q'}= \big\{\big(p_{\theta_u},\ket{\psi'_{\theta_u}}\big), \big(p_{\theta_d},\ket{\psi'_{\theta_d}}\big)\big\}$ (see Appendix), whose corresponding density matrix, $\rho'_{out}$, exhibits the average entanglement of $\bar{\cal Q}$.

We investigate two special cases. First we take $\theta=0$ (the measurement basis of $O$ is thus its natural basis), yielding the quantum ensemble $\bar{\cal Q}_1(p)=\{(1-p,\ket{\Psi^-}),(p,\ket{\Phi^-})\}$. For it we have $E_{{\cal Q}_1}(p)=1$ for any value of $p$. This entanglement is restored by the local control which flips the photon $B$ polarization each time the system is the state $\ket{\Phi^-}$. This produces the output state $\rho'_{out_1}=\ketbra{\Psi^-}{\Psi^-}$, whose concurrence $C_{cont}(p)=1$. We plot the corresponding entanglement obtained after the control loop as a function of $p$ in Fig.~\ref{Fig_results_closed}~a, see violet line (theory) and triangles (experiment): the restoration of the initial entanglement is achieved for any value of $p$ when the measurement basis angle is $\theta=0$.
As a second example, we consider the case in which the entanglement for the uncontrolled dynamics vanishes, which happens for $p=0.5$. 
We show that the amount of entanglement recovery depends on the measurement performed on $O$, that is to say on the angle $\theta$ of the rotated measurement basis. This measurement selects the system $AB$ quantum ensemble $\bar{\cal Q}_2(\theta)=\big\{\big(1/2,\ket{\varphi_{\theta_u}}\big), \big(1/2,\ket{\varphi_{\theta_d}}\big)\big\}$ (see Appendix). 
For it we find $E_{{\cal Q}_2}(\theta)=E_f(|\cos(2\theta)|)$, since the concurrence of both states $\ket{\varphi_{\theta_k}}$ is $|\cos(2\theta)|$. The $O$ measurement by itself does not produce any effect on the $AB$ entanglement, which is vanishing for any $\theta$ in that case, see black line (theory) and points (experiment) in Fig.~\ref{Fig_results_closed}~b. 
Instead application afterwards of the local control on the qubit $B$ may lead to recovery of the average entanglement of the ensemble, $\bar{\cal Q}_2(\theta)$. Indeed, the resulting output state for $AB$ is
\begin{equation}
   \rho'_{out_2}(\theta)=\cos^2\theta\ketbra{\Psi^-}{\Psi^-}\,+\,\sin^2\theta\ketbra{\Phi^-}{\Phi^-},
\label{eq-results:2ndExp-06}
\end{equation}
whose concurrence is $C_{cont}(\theta)=|\cos(2\theta)|$, see the pink line (theory) and triangles (experiment) in Fig.~\ref{Fig_results_closed}~b. Here we see that the natural basis of $O$ is the optimal measurement basis to recover the entanglement, whereas for $\theta=\frac{\pi}{4}$ no entanglement is regained.

The experimental data (symbols) plotted in Fig.~\ref{Fig_results_closed} show some deviation from the ideal expectations (continuous line). To explain them, we take into account the imperfect state preparation by our SPDC source: we also plot on Fig.~\ref{Fig_results_closed} the expected entanglement for a mixed input state whose fidelity with the ideal input state is $F_{\Psi^-}=0.90$ (dotted line) or $0.95$ (dashed line), see Appendix.

\section*{Conclusion}

In this article we have presented two experiments showing that recovery of entanglement after the interaction of the members of an entangled state with a local non-Markovian environment can be achieved on-demand through \emph{local} operations only. 
In the first experiment, the local environment is classical and induces low frequency noise, so that entanglement may be recovered by a local echo pulse. 
In the second experiment, the environment is a small quantum system (here a qubit) accessible to measurement, whose output controls a subsequent local operation on the system: as a result, entanglement recovery is obtained.

We note that in the second experiment there exists a measurement basis that selects system states with the largest amount of average entanglement. This maximum entanglement, recovered by the subsequent local operation, is known as entanglement of assistance~\cite{divincenzo,verstraete}, a concept of which we provide here an experimental illustration.
Instead the first experiment is a prototype of situations where the environment is not accessible in practice, typically because a large number of uncontrolled degrees of freedom are involved. Nevertheless, the recovery of entaglement by local operations is possible, since entanglement is not actually destroyed, but rather hidden due to a lack of classical information about which element in an ensemble of entangled states we are dealing with~\cite{darrigo2014Annals}.

Interestingly, the scheme of our first experiment is the simplest instance of recently proposed architectures, which should limit the detrimental up-scaling hardware-intrinsic decoherence, namely networks of smaller mutually entangled subsystems. Striking examples are trapped ions where modes in a single trap become dense as the trap becomes larger, acting as stray degrees of freedom, besides exquisitely solid-state systems as electrostatically defined quantum dots, silicon-based implanted impurities and optically active dopants~\cite{ladd2010}. Very effective on-demand recovery by local operations is possible when noise has strong low-frequency components. We stres that this is a very relevant physical case, $1/f$ noise being a major drawback for solid state quantum computation~\cite{PaladinoReview2013,ladd2010}, and it is likely to be an important source of decoherence in practical computation and communication networks. 
Direct applications of the scheme of our second experiment can be also envisaged in cavity quantum electrodynamics (QED)~\cite{haroche2006,ritter2012} or in circuit QED systems~\cite{wallraff2012}, where two-level quantum systems are strongly coupled to discrete photon modes in high-quality cavities. Finally we remark that this work, besides experimentally demonstrating novel physics related to non-Markovianity of open quantum systems, namely entanglement recovery by local operations, envisages new applications of quantum control techniques~\cite{ithier2005PRB,cook2007,bylander2011NatPhys,Vijay2012,Murch2013} to distributed architectures.

\section*{Acknowledgements}
G.B. acknowledges support from MIUR-PRIN 2011. P.M., F.S. and A.O. acknowledge support from the European project QWAD, Quantum Waveguide Application and Devices, http://www.qwad-project.eu/. R.L.F. acknowledges support by the Brazilian funding agency CAPES [Pesquisador Visitante Especial-Grant No. 108/2012]. A.d'A and R.L.F. acknowledge support from Centro Siciliano di Fisica Nucleare e Struttura delle Materia (Catania). 

\section*{Appendix}

\subsection{Entanglement of formation}
As entanglement measure $E_\rho$ we use the \emph{entanglement of formation} $E_f$~\cite{PlenioReview,HorodeckiReview}, which can be directly obtained from the concurrence $C(\rho)$~\cite{wootters1998PRL} by the following formula:
\begin{equation}
E_f(C)=\mathtt{h}\Big(\frac{1+\sqrt{1-C(\rho)^2}}{2}\Big), 
\label{methods:wootters-formula}
\end{equation}
where $\mathtt{h}(x)=-x\log_2x-(1-x)\log_2(1-x)$ and $C(\rho)=\max\{0,\lambda_1-\lambda_2-\lambda_3-\lambda_4\}$.
Here $\lambda_i$ ($\lambda_{i}\ge\lambda_{i+1}$) are the square roots of the eigenvalues of the non-Hermitian matrix $\rho (\sigma_y\otimes \sigma_y) \rho^\star (\sigma_y\otimes \sigma_y)$, $\rho^\star$ being the conjugate of $\rho$.

\subsection{Open-loop control}

\textbf{Random phases.--}
The random variables $\chi_k$ are Gaussian with a standard deviation $\sigma=0.6$ $rad$ and an average $\bar{\chi}\equiv\langle \chi_k \rangle=\frac{\pi}{2}$, $\langle \cdot \rangle$ being the ensemble average. Since we disposed of half-wave LCs, which could only generate phases in $[0,\pi]$, this setting provided that $99\%$ of the random phases were inside the experimentally achievable interval. Each sequence of random variables $\vec{\chi}=\{\chi_1,\chi_2,\chi_3,\chi_4\}$ were generated (using {\em Scilab} random number generation function "{\em grand}") by keeping $\chi_i$ equal to $\chi_{i-1}$ with probability $\mu$ and $\chi_i$ independent from $\chi_{i-1}$ with probability $1-\mu$.
 
\noindent The degree of memory in $\vec{\chi}$ may be quantified by $\mu$, which corresponds to the correlation coefficient~\cite{Papoulis1965} between $\chi_k$ and $\chi_{k+1}$ inside each sequence: 
\begin{equation}
\mu=\frac{\langle(\chi_k-\bar{\chi})(\chi_{k+1}-\bar{\chi})\rangle}{\sqrt{\langle(\chi_k-\bar{\chi})^2\rangle\langle(\chi_{k+1}-\bar{\chi})^2\rangle}}.
\label{methods:correlation-coefficient}
\end{equation}


\textbf{Imperfect initial state.--}
To take into account the effect of the imperfect state generated by our SPDC source, we consider the following partially mixed state as input sate:
\begin{eqnarray}
\rho_{in}=\eta\ketbra{\Psi^-}{\Psi^-}+(1-\eta)\frac{1}{4}\openone_4,
\label{methods:mixed-initial-state}
\end{eqnarray}
where input state $\frac{1}{4}\openone_4$ is the maximally mixed input state and the mixing parameter $\eta$ is related to the fidelity $F_{\Psi^-}=\bra{\Psi^-}\rho_{in}\ket{\Psi^-}$ by
\begin{equation}
\eta=\frac{4 F_{\Psi^-}-1}{3}.
\label{methods:fidelity-mixed-initial-state}
\end{equation}


\textbf{Theoretical calculation of the entanglement.--}
We calculate the output state of the two-photon system when the initial state is $\rho_{in}=\ketbra{\Psi^-}{\Psi^-}$. 

\noindent For the \textit{uncontrolled} dynamics, the two-photon state averaged on $\vec{\chi}$ is:
\begin{equation}
\rho_{out}(k)=\int d\vec{\chi} \,p(\vec{\chi})\, \ketbra{\bar{\nu}_{\chi_1,...,\chi_k}}{\bar{\nu}_{\chi_1,...,\chi_k}},
\label{eq-methods:1stExp_densityoperator}
\end{equation}
where $\ket{\bar{\nu}_{\chi_1,...,\chi_k}}$ is given by Eq.~(\ref{eq-results:1stExp_pure-state}).
We will use the following abbreviations: $\rho_{{ij}}(t_n)\equiv\bra{i}\rho_{}(t_n) \ket{j}$ where $i,j\, \in \{a,b,c,d\}$ and $\ket{a}\equiv\ket{00},\, \ket{b}\equiv\ket{01},\,\ket{c}\equiv\ket{10},\, \ket{d}\equiv\ket{11}$.
The noise only affects the coherences $\rho_{out_{bc}}$. These coherences are calculated by averaging the phase factor $e^{i\varphi_k}$:
\begin{equation}
\rho_{out_{bc}}(k)\,=\,\frac{1}{2} \Big\langle e^{-i\sum_{j=1}^k \chi_j}\Big\rangle,
\label{eq-methods:1stExp_densityoperator_bc}
\end{equation}
whose explicit expressions are:
\begin{eqnarray}
&& \hspace{-0.3cm}\rho_{out_{bc}}(1)=-\frac{1}{2}e^{-i\bar{\chi}}e^{-\frac{1}{2}\sigma^2},\label{eq:setup3-eq8}\nonumber\\
&& \hspace{-0.3cm}\rho_{out_{bc}}(2)=-\frac{1}{2}e^{-2i\bar{\chi}}e^{-2\sigma^2}\big[\mu+(1-\mu)e^{\sigma^2}\big],\nonumber\\
&& \hspace{-0.3cm}\rho_{out_{bc}}(3)=-\frac{1}{2}e^{-3i\bar{\chi}}e^{-\frac{9}{2}\sigma^2}
               \big[\mu^2+2\mu(1-\mu)e^{2\sigma^2}\nonumber\\
         &&\hspace{4.5cm}+(1-\mu)^2e^{3\sigma^2}\big],\nonumber\\
&& \hspace{-0.3cm}\rho_{out_{bc}}(4)=-\frac{1}{2}e^{-4i\bar{\chi}}e^{-8\sigma^2}
               \Big\{\mu^3+\mu^2(1-\mu)\big[2e^{3\sigma^2}+e^{4\sigma^2}\big]\nonumber\\
            &&\hspace{2.5cm}+3\mu(1-\mu)^2e^{5\sigma^2}+(1-\mu)^3e^{6\sigma^2}\Big\}.
\label{methods:exp-1-output-state-free-dynamics}
\end{eqnarray}

\noindent For the \textit{echoed dynamics}, we have to replace in Eq.~(\ref{eq-methods:1stExp_densityoperator}), for $k=3,4$, $\ket{\bar{\nu}_{\chi_1,...,\chi_k}}$ with $\ket{\tilde{\nu}_{\chi_1,...,\chi_k}}$ from the state (\ref{eq-results:1stExp_pure-state-echoed}). 
We call $\tilde{\rho}_{out}$ the new average state. The only non-trivial element we have to calculate is
\begin{equation}
\tilde{\rho}_{out_{ad}}(k)\,=\,\frac{1}{2} \Big\langle e^{-i\big(\chi_1+\chi_2-\sum_{j=3}^k \chi_j\big)}\Big\rangle,
\label{eq-methods:1stExp_densityoperator_ad}
\end{equation}
whose explicit expressions are:
\begin{eqnarray}
&& \hspace{-0.3cm}{\tilde{\rho}}_{out_{ad}}(3)=-\frac{1}{2}e^{-i\bar{\chi}}e^{-\frac{1}{2}\sigma^2}
               \Big\{\mu^2+\mu(1-\mu)\big[1+e^{-2\sigma^2}\big],\nonumber\\
&&\hspace{3.2cm}+\,(1-\mu)^2e^{-\sigma^2}\Big\},\label{eq:setup3-eq9}\nonumber\\
&& \hspace{-0.3cm} {\tilde{\rho}}_{out_{ad}}(4)=-\frac{1}{2}
               \Big\{\mu^3+\mu^2(1-\mu)\big[2e^{-\sigma^2}+e^{-4\sigma^2}\big]\nonumber\\
&&\hspace{2.2cm}\,+\,\mu(1-\mu)^2\big[2e^{-3\sigma^2}+e^{-\sigma^2}\big]\nonumber\\
&&\hspace{2.2cm}\,+\,(1-\mu)^3e^{-2\sigma^2}\Big\}.
\label{methods:exp-1-output-state-echoed-dynamics}
\end{eqnarray}

\noindent By means of the equations (\ref{methods:exp-1-output-state-free-dynamics}), (\ref{methods:exp-1-output-state-echoed-dynamics}) and by taking into account that the considered noisy channel is unital (the channel does not change the input state $\frac{1}{4}\openone_4$), it is straightforward to calculate the corresponding output state for the initial state (\ref{methods:mixed-initial-state}), simply by using the linearity of quantum operations.
  
\noindent The system $AB$ concurrence can be calculated by the formula
\begin{eqnarray}
&&\hspace{-0.5cm}C{(\rho(t))}=2\, \max\{0,\,|\rho_{{bc}}(t)|-\sqrt{\rho_{aa}\rho_{dd}},\nonumber\\
&&\hspace{2.6cm}   |\rho_{{ad}}(t)|-\sqrt{\rho_{bb}\rho_{cc}}\},
\end{eqnarray}
since the $AB$ density matrix assumes always a X-form~\cite{yu2007QIC}. In the case of \textit{uncontrolled} dynamics, the concurrence reads
\begin{equation}
C_{unco}(k)=2\, \max\Big\{0,\,\eta|\rho_{out_{b,c}}(k)|-\frac{1-\eta}{4}\Big\},
\label{methods:Concurrence-UncontrolledDynamics-ImperfectPreparation}
\end{equation}
with $k\in\,\{1,2,3,4\}$, whereas in the case of \textit{echoed} dynamics, the concurrence is given by
\begin{equation}
C_{echo}(k)=2\,\max\Big\{0,\,\eta|\tilde{\rho}_{out_{a,d}}(k)|-\frac{1-\eta}{4}\Big\},
\label{methods:Concurrence-EchoedDynamics-ImperfectPreparation}
\end{equation}
with $k\in\,\{3,4\}$. Obviously, for $\eta=1$ we obtain the formula relative to the perfect state preparation $\rho_{in}=\ketbra{\Psi^-}{\Psi^-}$. From (\ref{methods:Concurrence-UncontrolledDynamics-ImperfectPreparation}) and (\ref{methods:Concurrence-EchoedDynamics-ImperfectPreparation}), we can derive the entanglement of formation by means of Eq.~(\ref{methods:wootters-formula}).

\subsection{Closed-loop control}

\textbf{Actual experimental set-up.--} 
The actual set-up that was implemented for this second experiment is shown in Fig.~\ref{Fig_setup_closed}~c. For the sake of phase-stability and convenience, we opted for a folded version of the set-up shown in Fig.~\ref{Fig_setup_closed}~b: the two Mach-Zehnder interferometers were replaced by two Sagnac interferometers using a single beam-splitter. Likewise, we chose to measure the exit modes "$\theta_u$" and "$\theta_d$" on the same exit port "$\theta_{u/d}$": for a given angle $\theta$, "$\theta_u$" was measured for $\phi=2\theta$ and "$\theta_d$" was measured for $\phi=\pi-2\theta$.


\textbf{Ensembles of pure states.--} 
In the ensemble of pure states $\bar{\cal Q}$, 
$p_{\theta_u}=(1-p)\cos^2\theta+p\sin^2\theta$, $p_{\theta_d}=(1-p)\sin^2\theta+p\cos^2\theta$, $\ket{\psi_{\theta_u}}=(1-p)\cos\theta\ket{\Psi^-}-ip\sin\theta\ket{\Phi^-}$ and $\ket{\psi_{\theta_d}}=-i(1-p)\sin\theta\ket{\Psi^-}+p\cos\theta\ket{\Phi^-}$. With regard to the quantum ensemble
$\bar{\cal Q'}$, we have
$\ket{\psi'_{\theta_u}}=\ket{\psi_{\theta_u}}$ and $\ket{\psi'_{\theta_d}}=-i(1-p)\sin\theta\ket{\Phi^-}+p\cos\theta\ket{\Psi^-}$. 
For the ensemble $\bar{\cal Q}_2(\theta)$,
$\ket{\varphi_{\theta_u}}=\cos\theta\ket{\Psi^-}-i\sin\theta\ket{\Phi^-}$ and $\ket{\varphi_{\theta_d}}=-i\sin\theta\ket{\Psi^-}+\cos\theta\ket{\Phi^-}$.


\textbf{Imperfect state preparation.--} 
To take into account the imperfection of the state preparation (originating from the SPDC source and the CNOT wave plate mainly), we modeled the initial state as in Eq.~\ref{methods:mixed-initial-state}. The fidelity $F_{\Psi^-}$ of the initial $AB$ state was comprised in the interval $\{0.90,0.95\}$, it was estimated by measuring separately the fidelity of the $\ket{\Psi^-}$ and $\ket{\Phi^-}$ states in the "up" and "down" paths respectively. Note that in Fig.~\ref{Fig_results_closed}~a, the entanglement measured for $p=1$ is slightly lower than for $p=0$ (violet triangles) mainly because of the noise added by the CNOT on $\ket{\Phi^-}$.


\textbf{Theoretical calculation of the entanglement.--} 
For the \textit{uncontrolled} dynamics, when the input state is the mixed state given by (\ref{methods:mixed-initial-state}), the output state $\rho_{out}(p)$ has a concurrence:
\begin{eqnarray}
&&\hspace{-0.5cm} C(\rho_{out})=\frac{1}{2}\max\{0, \pm \,2\eta (1-2p)-1+\eta\}.
\label{methods:mixed-output-state-concurrence-unco}
\end{eqnarray}

\noindent For the \textit{controlled} dynamics and the same input state, the output state has the concurrence:
\begin{eqnarray}
&&\hspace{-0.5cm} C(\rho'_{out})=\frac{1}{2}\max\{0,\eta(1\pm2\cos2\theta)-1\}.
\label{methods:exp2-controlled-mixed-output-state-concurrence}
\end{eqnarray}


\textbf{Error bars.--} 
The errors bars on Fig.~\ref{Fig_results_closed} are calculated from the Poissonian statistical errors associated to the coincidence counts.

\noindent The vertical error bars stem from the propagation of the Poissonian statistical errors of the 36 coincidence counts used for the quantum tomography that allows to measure $E_f$. 

\noindent The horizontal error bars, for the measurements done varying $p'$ (Fig.~\ref{Fig_results_closed}~a), stem from the Poissonian statistical errors of the coincidence counts used to set $p'$:
\begin{equation}
p' = \frac{C_{HHd}+C_{VVd}}{C_{HVu}+C_{VHu}}, \nonumber
\end{equation}
where $C_{HHd}$ and $C_{VVd}$ ($C_{HVu}$ and $C_{VHu}$) are measured coincidence counts corresponding to the state $\ket{\Phi^-}$ on the path $\ket{d}$ (to the state $\ket{\Psi^-}$ on the path $\ket{u}$).
The error $\delta p'$ on $p'$ is thus given by:
\begin{eqnarray}
\delta p'^2 &=& \sum_j \Big(\frac{\partial p'}{\partial C_{j}}\Big)^2.(\delta C_{j})^2\,\textrm{, }j=HHd, HVu, VVd, VHu \nonumber\\
&=& \frac{C_{HHd}}{(C_{HVu}+C_{VHu})^2} + \frac{C_{HVu}.(C_{HHd}+C_{VVd})^2}{(C_{HVu}+C_{VHu})^4} \nonumber\\
&+& \frac{C_{VVd}}{(C_{HVu}+C_{VHu})^2} + \frac{C_{VHu}.(C_{HHd}+C_{VVd})^2}{(C_{HVu}+C_{VHu})^4}, \nonumber
\end{eqnarray}
where $\delta C_{j}$ is the the Poissonian statistical error on $C_j$.

\noindent The horizontal error bars, for the measurements done varying $\theta$ (Fig.~\ref{Fig_results_closed}~b), stem from the Poissonian statistical errors of the coincidence counts used to set $\theta$:
\begin{equation}
\theta = \arctan(\sqrt{R}) \nonumber\,\textrm{, with}\,R = \frac{C_{HV1}+C_{VH1}}{C_{HV0}+C_{VH0}}, \nonumber
\end{equation}
where $C_{HV1}$ and $C_{VH1}$ ($C_{HV0}$ and $C_{VH0}$) are the coincidence count values measured on the output mode $\theta_{d}$ ($\theta_{u}$).
The error $\delta \theta$ on $\theta$ is given by:
\begin{eqnarray}
\delta \theta^2 &=& \sum_j \Big(\frac{\partial \theta}{\partial C_{j}}\Big)^2.(\delta C_{j})^2\,\textrm{, }j=HV1, VH1, HV0, VH0 \nonumber\\
&=& \frac{1}{2} \frac{1}{\sqrt{C_{HV0}+C_{VH0}}}\frac{1}{\sqrt{1+R}}. \nonumber
\end{eqnarray}

\bibliography{bibHE}

\end{document}